\documentclass[sigconf]{acmart}

\AtBeginDocument{%
  \providecommand\BibTeX{{%
    \normalfont B\kern-0.5em{\scshape i\kern-0.25em b}\kern-0.8em\TeX}}}

\usepackage{graphicx,wrapfig}
\usepackage{subfigure}
\usepackage{enumitem}
\usepackage{pifont}
\usepackage{multirow}
\usepackage[ruled,linesnumbered]{algorithm2e}
\usepackage{caption}
\usepackage{graphicx}
\copyrightyear{2021}
\acmYear{2021}
\setcopyright{acmcopyright}
\acmConference[BSCI '21] {Proceedings of the 3rd ACM International Symposium on Blockchain and Secure Critical Infrastructure}{June 7, 2021}{Virtual Event, Hong Kong.}
\acmBooktitle{Proceedings of the 3rd ACM International Symposium on Blockchain and Secure Critical Infrastructure (BSCI '21), June 7, 2021, Virtual Event, Hong Kong}
\acmPrice{15.00}
\acmISBN{978-1-4503-8400-1/21/06}
\acmDOI{10.1145/3457337.3457846}

\begin{document}
\fancyhead{}

\title{An Anomaly Event Detection Method Based on GNN Algorithm for Multi-data Sources}

\fancyhead{}
\author{Yipeng Ji$^{1}$, Jingyi Wang$^{1}$, Shaoning Li$^{2}$, Yangyang Li$^3$, Shenwen Lin$^{4}$, Xiong Li$^{5,6}$}
\affiliation{
\institution{
$^1$School of Computer Science and Engineering, Beihang University, Beijing, China;\\ 
$^2$School of Cyber Science and Technology, Beihang University, Beijing, China;\\ 
$^3$National Engineering Laboratory for Public Safety Risk Perception and Control by Big Data, Beijing, China;\\
$^4$National Computer Network Emergency Response Technical Team of China (CNCERT), Beijing, China;\\
$^5$Beijing Biwei Network Technology Co., Ltd., Beijing, China; 
$^6$Beijing Xindatek Technology Co., Ltd. Beijing \country{China};\\
}}
\email{jiyipeng,wjy10373,19373215@buaa.edu.cn; lsw@cert.org.cn; liyangyang@cetc.com.cn; li.xiong@foxmail.com}
\authornote{Jingyi Wang is the corresponding author.}

\begin{abstract}
 Anomaly event detection is crucial for critical infrastructure security(transportation system, social-ecological sector, insurance service, government sector etc.) due to its ability to reveal and address the potential cyber-threats in advance by analysing the data(messages, microblogs, logs etc.) from digital systems and networks. 
 However, the convenience and applicability of smart devices and the maturity of connected technology make the social anomaly events data multi-source and dynamic, which result in the inadaptability for multi-source data detection and thus affect the critical infrastructure security. 
 To effectively address the proposed problems, in this paper, we design a novel anomaly detection method based on multi-source data.
 First, we leverage spectral clustering algorithm for feature extraction and fusion of multiple data sources. 
 Second, by harnessing the power of deep graph neural network(Deep-GNN), we perform a fine-gained anomaly social event detection, revealing the threatening events and guarantee the critical infrastructure security. 
 Experimental results demonstrate that our framework outperforms other baseline anomaly event detection methods and shows high tracking accuracy, strong robustness and stability. 
\end{abstract}



\maketitle

\section{Introduction}
Secure critical infrastructure, such as transportation system, economic sector, insurance service, government sector etc., is essential for the functioning of a society and economy.
As smart devices and intelligent terminal advanced, critical infrastructures and their security increasingly come to rely on embedded control systems and external environment, especially social networks, which combined with massive hidden information on potential security threats \cite{zhao2021automatically, liu2020alleviating}.
Due to the nature of events in social networks that is able to indirectly reflect such latent problems, detecting social events has become a research hotspot in data mining as well as security field to solve the attacks in advance. 

Social abnormal event is a phenomenon that exists in a large number of social networks\cite{peng2021lime, mao2021event, peng2019fine}, and these abnormal event contain a lot of hidden information related to reality; by in-depth analysis and timely detection, it can realize effective management of social environment. Effective and timely detection is considered as an attractive method to discover the anomalies hidden or hot topic in the complex and ever-changing of multi-source data world, but there are many challenges to implement this approach; the challenges mainly include the variety of data types, the complex data sources differentiation and dependence complexity in multi-source data \cite{dou2020enhancing, liu2020event}. From theory, since anomaly detection has important theoretical and practical significance, they have an attractive research field; From the reality, anomaly detection research is closely related to our economic life and has great practical significance and economic value. 

Some model and method have been used in related field, and gain a lot of useful results. The most frequently used the methods and models of social anomaly detection include the followed several types: utilizing similar distance calculation method, utilizing similar density aggregation, clustering analytical method, high-dimensional characteristic and graph network to realized anomaly detection in multi-data source. In \cite{liu2020event,cao2021knowledge}, the paper considers the idea of the anomaly detection based on similar distance analytical method\cite{liu2021heterogeneous} is an hypothesis; the hypothesis is that the distance between the abnormal samples and the normal samples is larger than the distance from the normal sample to the data set; the anomaly degree is defined by the distance between the sample and other objects in the data set, and the method needs set a distance hyper-parameter and a ratio hyper-parameter to renew all the samples in the data set after a process of filter operation. The disadvantage of this method can only label binary samples in the data set and cannot useful for samples of the multi-data source directly, and the similar distance have  major limitations\cite{zhao2020multi}. 

Cluster-based method is a machine learning technique that involves grouping data points; the detailed description of the method is that given a set of data points, clustering algorithms can be used to divide each data point into a specific group\cite{hu2020toposch}. In theory, data points in the same group should have similar properties, and data points in different groups should have highly different attributes. 
Density-Based Spatial Clustering of Applications with Noise (DBSCAN) algorithm is an representative algorithm of anomaly detection methods based on clustering. 
However, the disadvantage of the method based on density is that the effectiveness of the model. The type of algorithm is heavily dependent on the effectively characterize the clustering structure of the specified data set, and the generality of the algorithm is ignored. 

With the popularity of intelligent communication equipment and the development of mobile Internet technology, social networking is becoming a tool that people use frequently; it forms a very large digital world of social behavior, and there are complex dependencies and correlation between the data of the diverse sources. The abnormal events of the social network can reflect the real emotions and opinions of network users, and becomes one of the important goal of big data mining in the network information space.
The primary task of event detection is how to accurately represent the characteristics of social network data, and can timely detect and analyze events in streaming online social data. Aforesaid method can not be effectively applied to detect social anomaly events in multi-source data, but graph structures can be accurately used to represent the interdependencies between objects and to build model. Graph-based anomaly detection methods have become the attention of more and more researchers, especially 
in multi-source data domain. Multi-source exception detection method has some advantages, and the methods used samples between different data sources to detect exceptions in a multi-source data set. 
In these methods, the anomaly object was defined as if a sample is assigned to different clusters in different data block. In literature \cite{2013Clustering}, the author considered the relationship between the sample in the specified data block and the possible clustering structures in that data block, to obtain the final anomaly score for each sample. 

Although existing methods of the social anomaly event have produced great effect in multi-data source, with the continuous expansion of the application field and the increasing complexity of the types of data sources, there remains some problem need to be solved further such as the incremental representation of data, feature fusion learning and emergency detection in multi-data source\cite{peng2017incrementally}. With the increasing of deep learning technology in social anomaly event detection, the model and  algorithm can bring new idea to address the practical problem. Since the dynamics, diversity and sparsity of social multi-data source cause difficult representation, poor fusion learning of model and low efficiency of event detection in multi-data source, the paper proposed a novel social anomaly event detection, and the method uses spectral clustering to realized feature extraction based on multi-data source, and uses deep graph neural network to accomplish the social anomaly event detection.

The rest of this paper is arranged as follows: Sections II mainly introduces the related work and algorithms, Sections III describes the model designed. The results of the experiment are show in Sections IV and the conclusion in Sections V.

\section{Related Works and Algorithms}
This section mainly reviews the related theories and introduces the useful algorithms in our paper.  

\subsection{The Spectral Clustering}

Spectral clustering is a widely used clustering algorithm, and by comparing to the traditional k-means algorithm, spectral clustering has stronger adaptability to realize data distribution, and has a good clustering effect and low computation. Spectral clustering is a clustering method based on graph theory, by clustering for the eigenvectors of Laplace matrix of sample data to achieve the purpose of sample data clustering. The spectral clustering algorithm is simply described as 
\begin{itemize}
\item N sample points is inputted $X=\{x_1,x_2,\ldots,x_n\}$ and the number of cluster clusters k; output clustering cluster $A_1,A_2,\ldots,A_k $.
\item The similarity matrix W of $n*n$ is calculated as 
the following formula:

\begin{equation}
s_{ij}=s(x_i,x_j)=\sum_{i=1,j=1}^n exp\dfrac{{-\lVert x_i-x_j\rVert}^2}{2{\sigma}^2}
\end{equation}
where W is similarity matrix of $s_{ij}$.
\item The degree matrix D is calculated as the following formula:
\begin{equation}
d_i=\sum_{j=1}^n{w_{ij}}
\end{equation}
where D is the diagonal matrix of $n*n$;   
\item The Laplace matrix is calculated $L=D-W$;
\item To calculate the eigenvalues of L, and sort the eigenvalues from smallest to largest, and extract the first k eigenvalues, and then calculate the eigenvectors of the first k eigenvalues $u_1,u_2,\ldots,u_k,$;
\item Take k of the above column vectors and form them into a matrix $U=\{u_1,u_2,\ldots,u_k\}$, $U\in R^{n*k}$;
\item Assum $y_i\in R^k$ is the $i^\textup{th}$ row vectors of U, where $i=1,2,\ldots,n$; 
\item Use k-means algorithm to place new sample points $Y=\{y1,y_2,\ldots,y_n$ to realize the cluster cluster $C_1,C_2,\ldots C_K$;
\item Output cluster $A_1,A_2,\ldots,A_k$, where, $A_i=\{j|y_j\in C_i\}$.
\end{itemize}

\subsection{Typical Algorithms}

According to the partition criterion used in the spectral clustering algorithm, the algorithm can be divided into two-way spectral clustering algorithm and multi-way spectral clustering algorithm\cite{Yu2003Multiclass}. The former uses two-way division criterion and the latter uses K-way division criterion. Two-way spectral clustering algorithm includes PF algorithm, SM algorithm, SLH algorithm, KVV algorithm, and Mcut algorithm. PF algorithm proposes the eigenvectors corresponding to the maximum eigenvalue of the similarity matrix $W$ are used for clustering, and points out that for block diagonal similarity matrices, the points corresponding to non-zero value of eigenvectors belong to the same class, and the points corresponding to zero value belong to another class\cite{White2005A}. SM algorithm points out that the difference between NCUT and MNCUT lies only in the spectral mapping used, and the eigenvector corresponding to the second smallest eigenvalue, namely Fiedler vector, contains the partition information of the graph, and according to the heuristic rule, the partition point $i$ in this vector is found to minimize the value of NCUT (A,B) obtained at this point, Finally, the value in the vector is compared with the minimum value of NCUT criterion function\cite{Rohe2012SPECTRAL}. The points are greater than this value or equal to this value, and the points belong to one class, and points are less than this value belong to another class. The SLH algorithm reposition algorithm calculates for the first $k$ feature vectors of the similarity matrix $W$, and the parameter $k$ needs to be specified in advance. KVV algorithm, according to the heuristic rules, the division point $i$ in the Fiedler vector is found to minimize the RCUT (A,B) value at this point, which is similar to the SM algorithm\cite{Challa2020Power}. The only difference is that the SM algorithm looks for the partition point that minimizes the value of NCUT (A,B).

Multi-path spectral clustering algorithm includes NJW algorithm and MS algorithm \cite{li2020modeling}. NJW algorithm, the eigenvectors corresponding to the first $k$ maximum eigenvalues of the Laplace matrix are selected to construct a new vector space $R$, and the corresponding relationship with the original data is established in this new space, and then the clustering is carried out\cite{cao2021knowledge}. In MS algorithm, the first $k$ non-zero eigenvalues of the random walk matrix $P$ are used to construct the matrix, and then the rows in the matrix are treated as points in $R$ space for clustering. The steps are similar to the NJW algorithm. The MS algorithm has achieved good results in the actual image segmentation, but the clustering effect will be poor if there is a big difference between the diagonal element values in the degree matrix $D$\cite{2020Spectral}.

\subsection{Graph Neural Network}

Recent years, graph neural network(GNN)has received more and more attention in many research field, including social networking, knowledge graphs, recommendation systems, etc.\cite{sun2020pairwise,peng2019hierarchical}. The powerful ability of GNN is to model the dependencies between nodes in graphs, and has made breakthroughs in the research fields related to graph analysis.
A basic idea of graph neural network is that the node is embedding based on its local neighbor information\cite{DAVID1999Toward,peng2020motif}. GNN is a kind of neural network which acts directly on graph structure. GNN has the following characteristics: the input order of nodes are ignore; in the process of calculation, the representation of a node is affected by its neighbor nodes, while the graph itself is connected unchanged; The representation of graph structure enables graph-based reasoning. Generally speaking, it is to aggregate the information of each node and its surrounding nodes through a neural network, Fig. 1 show the process of embedding:

\begin{figure}[htbp]
\centering\includegraphics[width=0.7\linewidth]{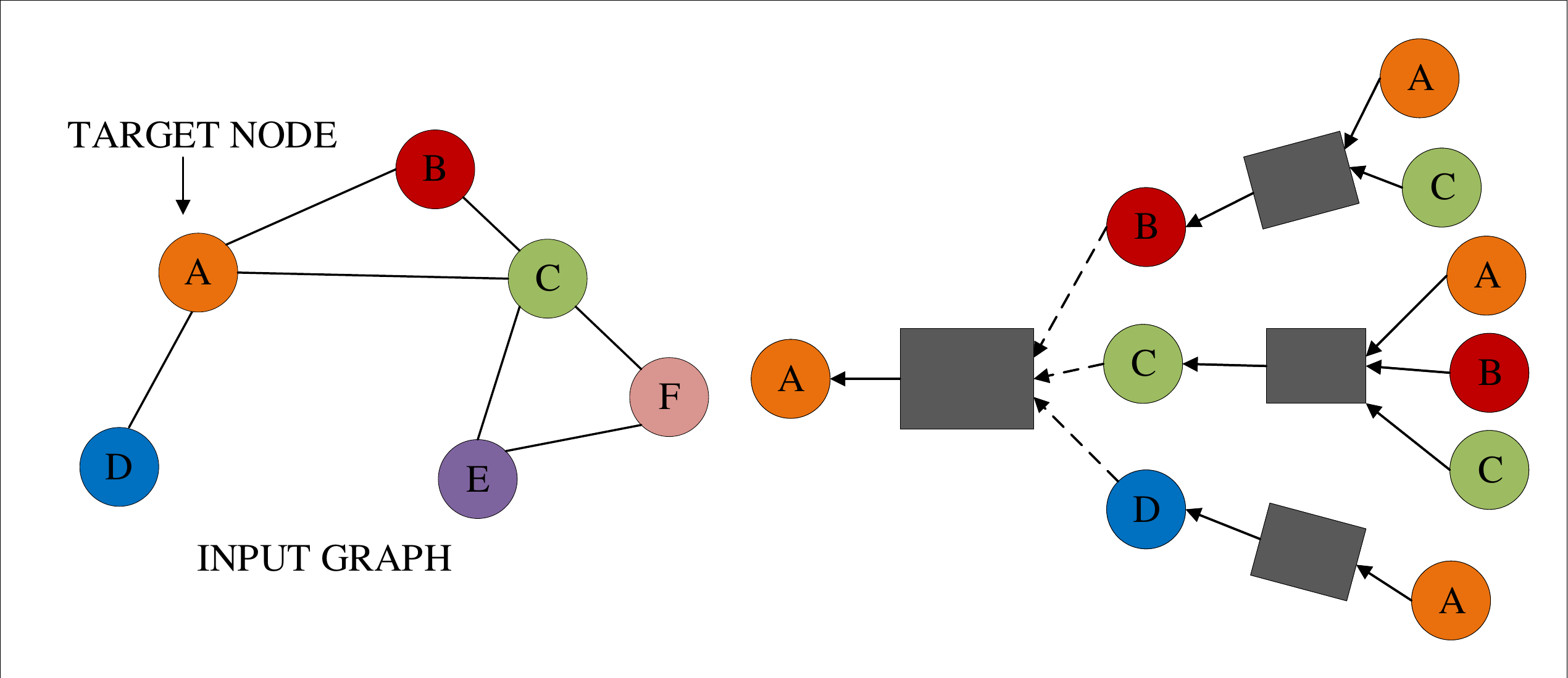}
\caption{The process of embedding.}
\end{figure}

Neural network is usually composed of two modules: propagation module and output module. Fig. 2 show the describes the basic structure of neural network:

\begin{figure}[htbp]
\centering\includegraphics[width=0.7\linewidth]{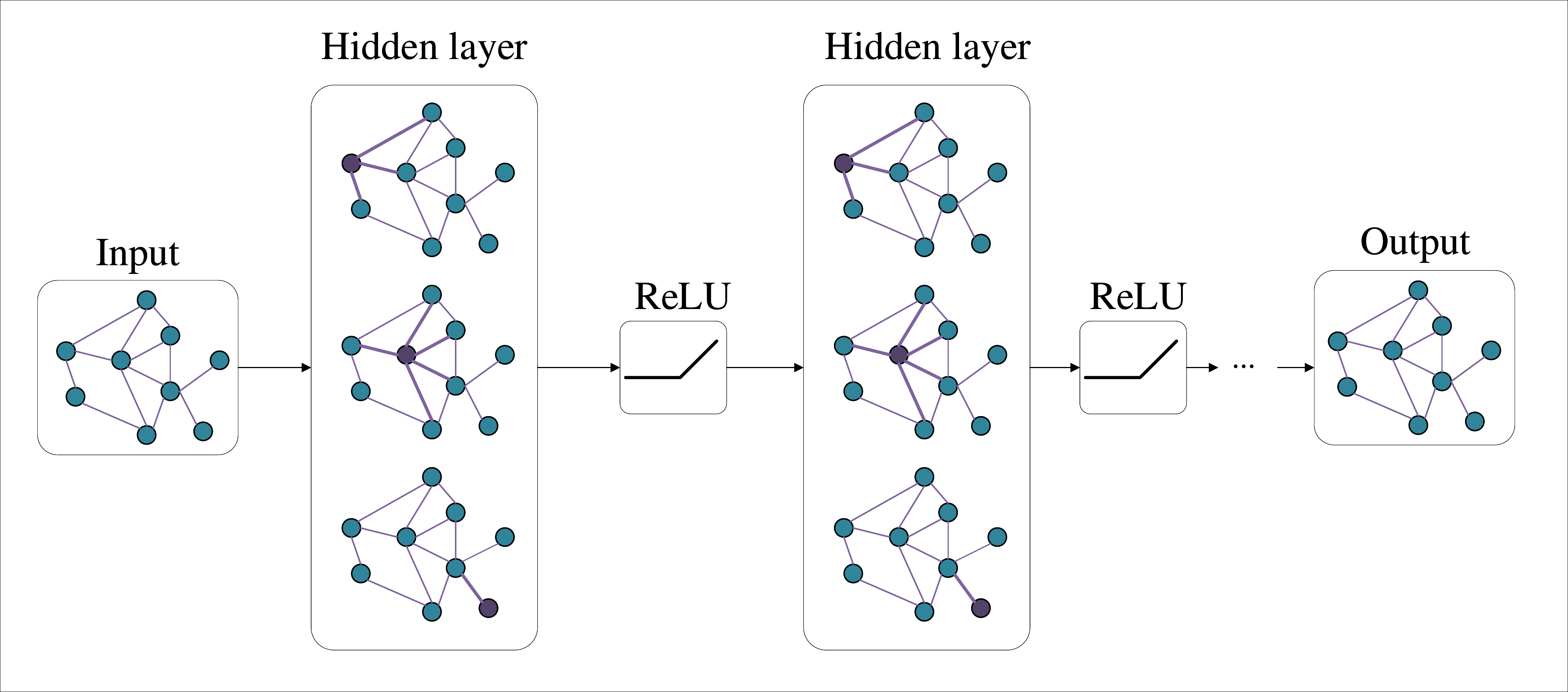}
\caption{The basic structure of neural network.}
\end{figure}

The symbols and meanings involved in the neural network structure as follow: $G=(N,E)$, N is the set of vertices, E is the set of edges; $ne[n]$ is the adjacency vertex of vertex $n$; $co[n]$ represents the edge attached to the vertex n;$x_n$ represents the characteristics of vertex n; $x_co[n]$ represents the characteristics of the edges attached to the vertex n; $h_ne[n]$ represents the embedded representation of vertex n (state embedding); $x_ne[n]$ represents the characteristics of adjacent nodes of vertex n; $h_n$ is the embedded representation of vertex n\cite{Asai2001Modelling}.

In propagation module of the GNN, the information is passed between nodes, and the status is updated, where there include aggregator and updater\cite{Deng2020Cola}. Aggregator: its purpose is to aggregate the information of the nodes around the node $n$, and to learn the embedding representation of node $n$, $h_n$, and $h_n$ is represented as
\begin{equation}
h_n=f(x_n,x_{co[n]},h_{ne[n]},x_{ne[n]}) 
\end{equation}
where $f$ can be interpreted as a feedforward fully connected neural network.
Updater: the embedded representation of node $N$ in during the learning process the model iteratively updates is the following formula 
\begin{equation}
H^{t+1}=F(H^t,X)
\end{equation}
where, $t$ is the iteration of $t^\textup{th}$, $X$ is all the features, and $H^t$ is the embedded representation of all nodes in the $t^\textup{th}$ iteration.

The output module of the GNN can be divided into two forms according to the different application types of the model:node-focused and graph-focused\cite{2020Point}. For node-focused classification, the label corresponding to each node after the last iteration is outputted by the model, \begin{equation}
o_n=g(h_n^T,x_n)
\end{equation}
where $t_i$ is the true label of the $i^\textup{th}$ node, and $o_i$ is the model output label, and $p$ is the number of the node. For the graph-focused classification, the READOUT function is used to aggregate node features from the last iteration, and to obtain the representation vector $h(G)$ of the entire graph.
\begin{equation}
h(G)=READOUT({h_v^{K}|v\in G})
\end{equation}
The READOUT function can be a simple permutation invariant function

\section{Model Design}
In this section, we mainly proposed an novel method of social anomaly event detection  based on multi-source data; the method used spectral clustering algorithm to realize the feature space extension in multi-source data, and utilizes GNN model to achieve the anomaly event detection. Fig. 3 shows the design process of social anomaly detection based on multiple data sources.

\begin{figure}[htbp]
\centering\includegraphics[width=0.8\linewidth]{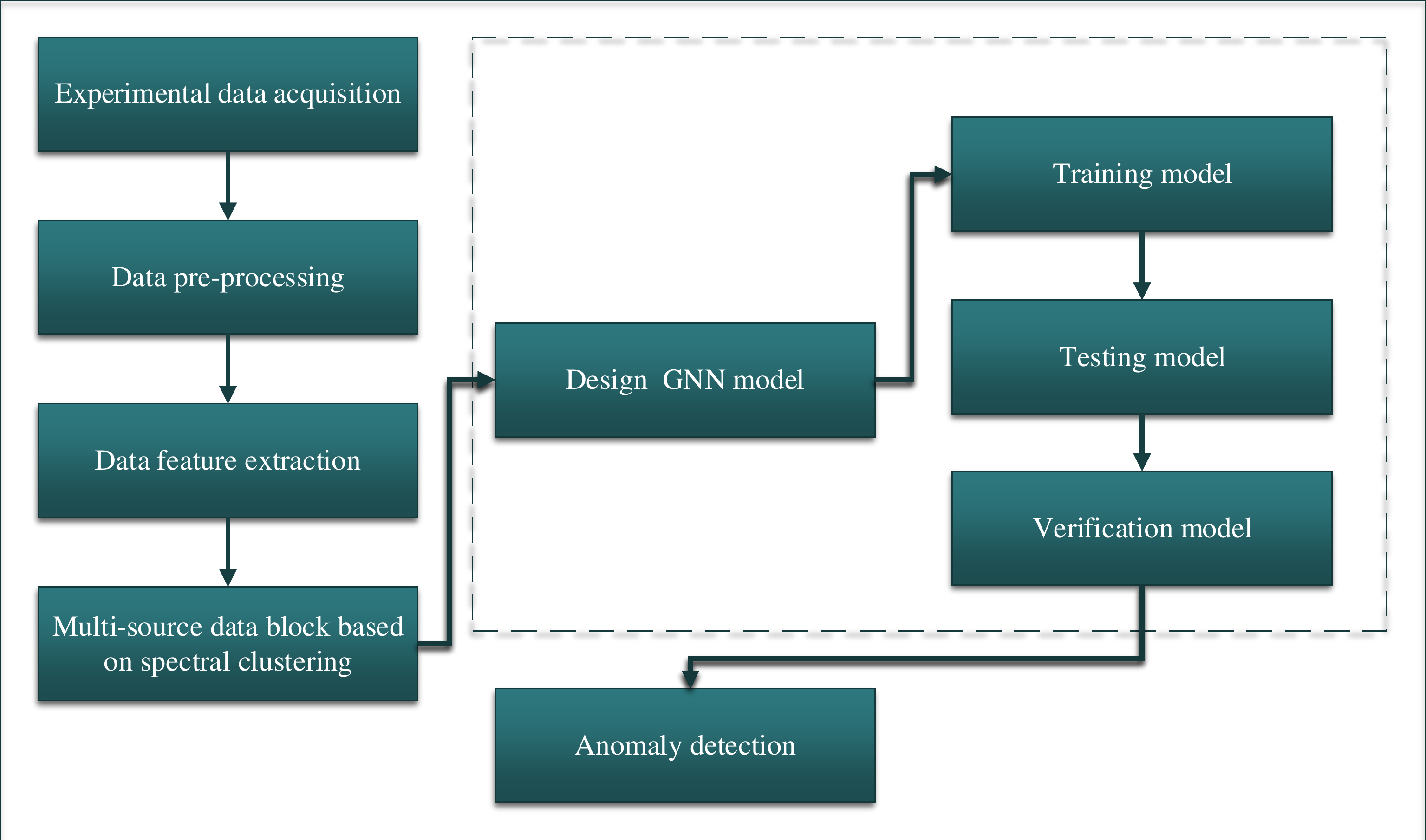}
\caption{Design process of social anomaly detection.}
\end{figure}

The processes of abnormal event detectionin this paper include the below the steps shown:
\begin{itemize}
\item Data collection stage collect data samples related to the detection requirements;
\item Data pre-processing includes data denoising and dimensionality reduction;
\item Model design, according to the detection requirements  the appropriate anomaly detection model is selected.
Set the necessary parameters for the model.
\item Model training is carried out and use data pre-processing to train model with supervised or unsupervised way, and this is the key step.
\item In model testing, the result of testing is acquired by anomaly detection model with the test data set.
\item Anomaly detection stage: the obtained model is used to detect anomalies in the target data.
\end{itemize}
\subsection{Feature Space Based on Spectral Clustering }
In this paper, the data generated from different data sources are analyze as a single view, and the multiple data are considered as multiple view learning. Samples from different data sources have different clustering forms, and the extended data set from multiple sources data can be considered that a description of the same problem from different angles\cite{2007A}. In multiple source data sets, anomaly event is redefined as the sample is of  inconsistent behavior in different data source , and the anomaly event of multiple view learning is defined as the sample belongs to the different clustering structure formed in different view spare learning\cite{Savage2014Anomaly}. 

We propose a consistent multi-data block anomaly detection method based on spectral clustering algorithm. In the method, the corresponding feature spaces of different data source are fused into an overall feature space, and using abnormal score to achieve anomaly judgement; in the data set, the abnormal score of the sample is defined by the each sample in different data source for the difference between the membership degree of each cluster structure\cite{2016Detecting}. 

In this paper, some important concepts are defined: the feature space of the different data source is called single feature space, the expression is $SF_t^{m\times d_k}=[A_t\in P^{m\times d_k}]$, where $A_t$ represents the data of $t^{\textup{th}}$ data source, $d_k$ represents dimension; the feature space of each data source constitutes a full feature space, and the expression is $CSF^{m\times (\sum_{k=1}^nd_k)}=[CSF_1^{m\times d_1},CSF_1^{m\times d_2},\ldots,CSF_k^{m\times d_k}]$. The difference between membership degree of anomaly data in full feature space is bigger than it in single feature space\cite{2013Information}. 

\subsection{Social Anomaly Detection based on GNN}
By spectral clustering, a full feature space of anomaly event based on multi data source and the characterization coefficient matrix is defined. First the model uses GNN to extracts the network element information (node, edge) of the full feature space at $t$, to form characterization of the feature, and then using the unsupervised representation learning algorithm DGI of the graph take the entire network at current moment to represent as a one-dimensional vector; Using RRCF algorithms to obtain the abnormal score of every moment\cite{Aggarwal2013Outlier}. In order to effectively use the graph information at every moment, we use long short term memory network (LSTMs) to obtain and process the change information of the global network representation at every moment. The framework of GNN algorithm is show in Fig. 4:

\begin{figure}[htbp]
\centering\includegraphics[width=0.8\linewidth]{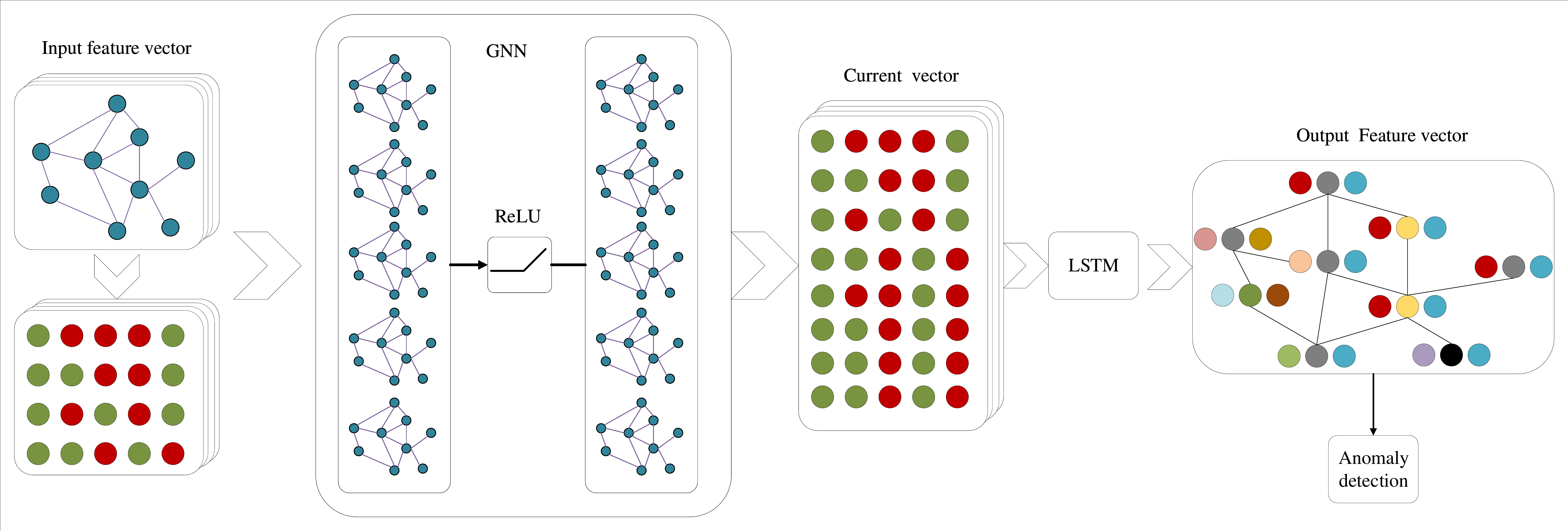}
\caption{Framework of GNN algorithm.}
\end{figure}

\section{Results and Discussions}
In this section, social network structure data and social text data collected online will be respectively used to evaluate the proposed method of this paper. 
\subsection{Experimental Data}\label{AA}

In the paper, we use real multi-sources data set for experiment, and the data set is main from the micro-blog data set and sina social platform data set. Twenty thousand social message records are randomly sampled from the collected social micro-blog data set and sina social platform; the data set includes text data, picture data, and video data; In the datasets the $60\%$ as a training sample, $20\%$ as verification  sample, and $20\%$ as test sample; the 10,000 social micro-blog messages contain 7,880 manually marked messages of social events. 

In order to eliminate the inaccurate data in the data set, and some irrelevant data are made by the network node exception, and some data collected repeatedly, and some data with wrong format or null value; these data need to be pre-processed to find out the data set; the data pre-processed is key step for mode, and feature representation provides reliable data guarantee for the algorithm\cite{2013Systematic}. In the paper, we use principal component analysis(PCA) as he method of the data pre-processed. PCA is a multivariate statistical analysis method in which several variables are linearly transformed to select a small number of important variables. Also it is known as principal component analysis. The idea of PCA is to delete the redundant variables (closely related variables) for all the variables originally proposed, and establish as few new variables as possible, so that these new variables are paredly unrelated, and these new variables keep the original information as much as possible in reflecting the information of the subject. The main steps of principal component analysis are as follows:
\begin{itemize}
\item Standardization of index data;
\item The correlation between the indicators;
\item Determine the number of principal components $m$;
\item Principal component expression;
\item Principal component naming.
\end{itemize}

\subsection{Experimental Result}\label{AA}

To better verify the SP-GNN method we mentioned in this paper, the other methods Support Vector Machine(SVM) and convolutional neural network (CNN) are employed as a contrast. Accuracy and ${\rm{Macro}}-F_1$ scores are selected as evaluation index of the model. In order to ensure the accuracy of the results, we repeated the experiment for three times and took the average value as the final result. The 
experimental results are shown in the Fig. 5.

\begin{figure}[htbp]
\centering\includegraphics[width=0.8\linewidth]{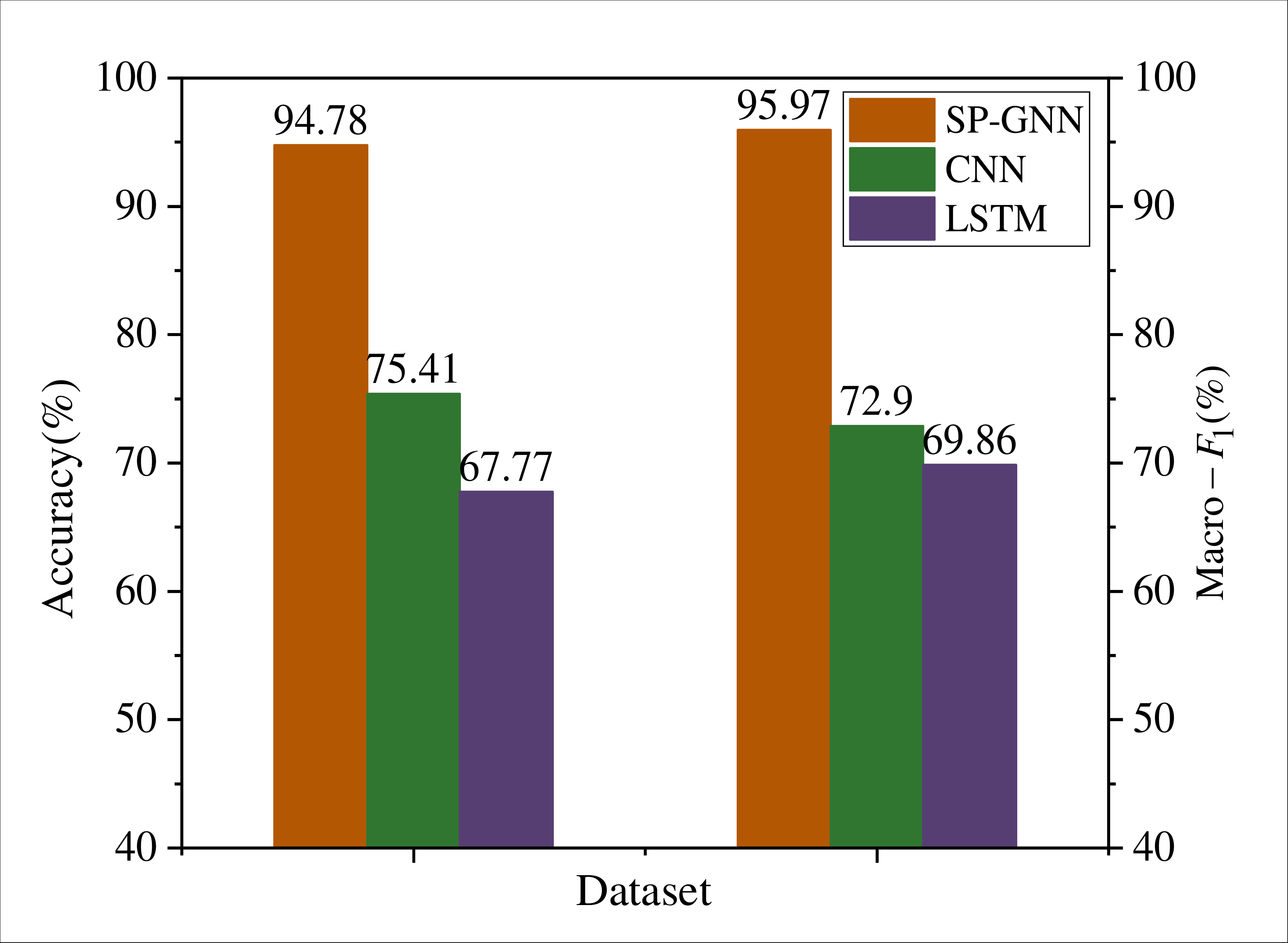}
\caption{Comparison of experimental results.}
\end{figure}

Accuracy and ${\rm{Macro}}-F_1$ of the SP-GNN model  in social anomaly events is respectively better than the other method. The experimental results fully illustrate the proposed method is an effective means in in social anomaly events detection based on multi-data source. 
By drawing Receiver Operating Characteristic(ROC) curve to analyse the performance of the proposed method and baseline methds. The ROC curve is the corresponding values for each result of model anomaly. The results are demonstrated in the Fig. 6.

\begin{figure}[htbp]
\centering\includegraphics[width=0.8\linewidth]{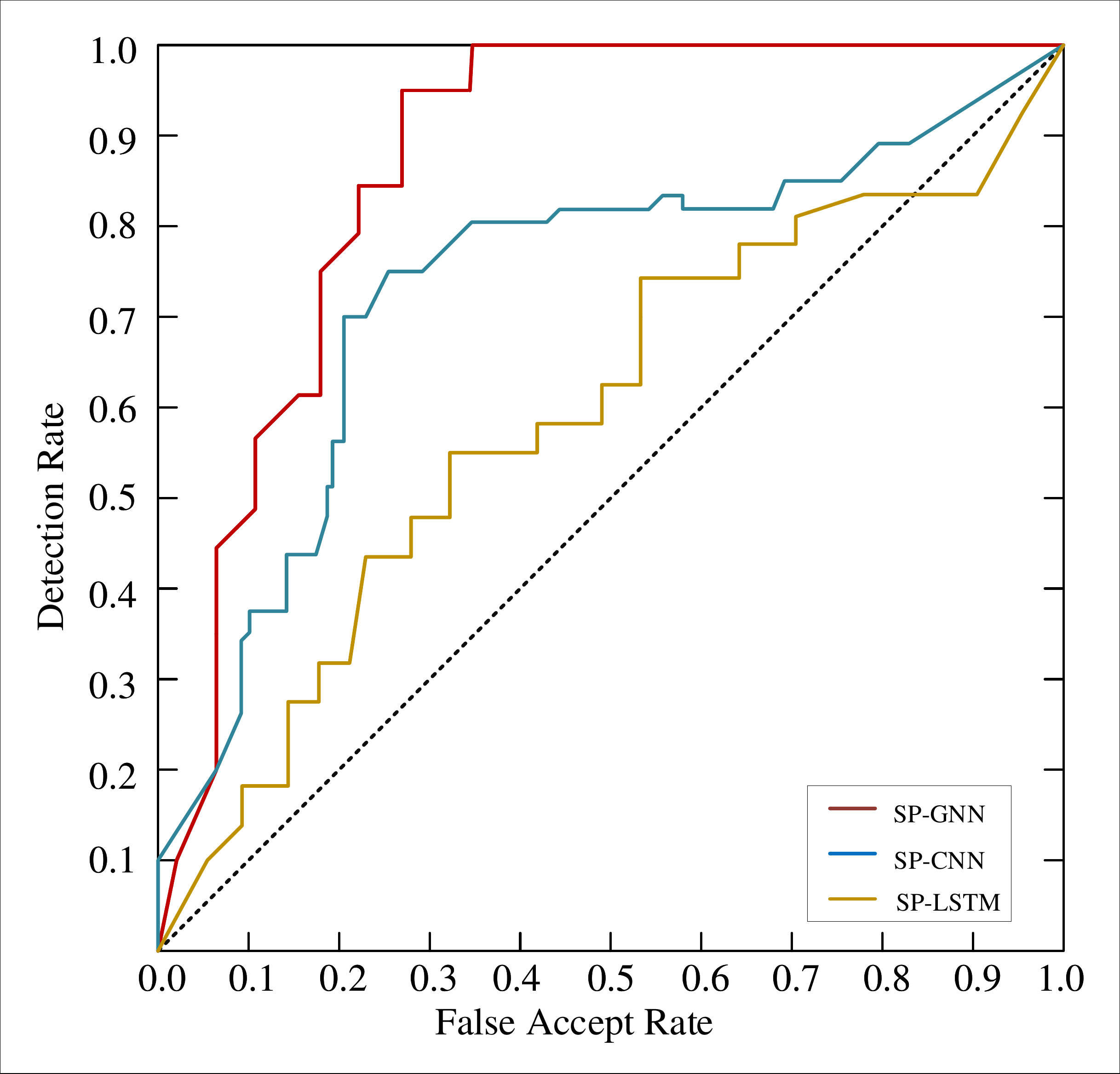}
\caption{The ROC curve for results of anomaly detection.}
\end{figure}
From the result, we can see that the area under the ROC curve is more bigger, and the performance of the model is better. The area under SP-GNN curve is the most biggest, and the performance of the SP-GNN method is superior to the baseline methods in exception object detection based on multi-data source.
As can be seen from Fig. 7, SP-GNN generally gives a high score to the anomaly graphs in the 800 to 1400 plots where the anomaly is most concentrated.
\begin{figure}[htbp]
\centering\includegraphics[width=0.8\linewidth]{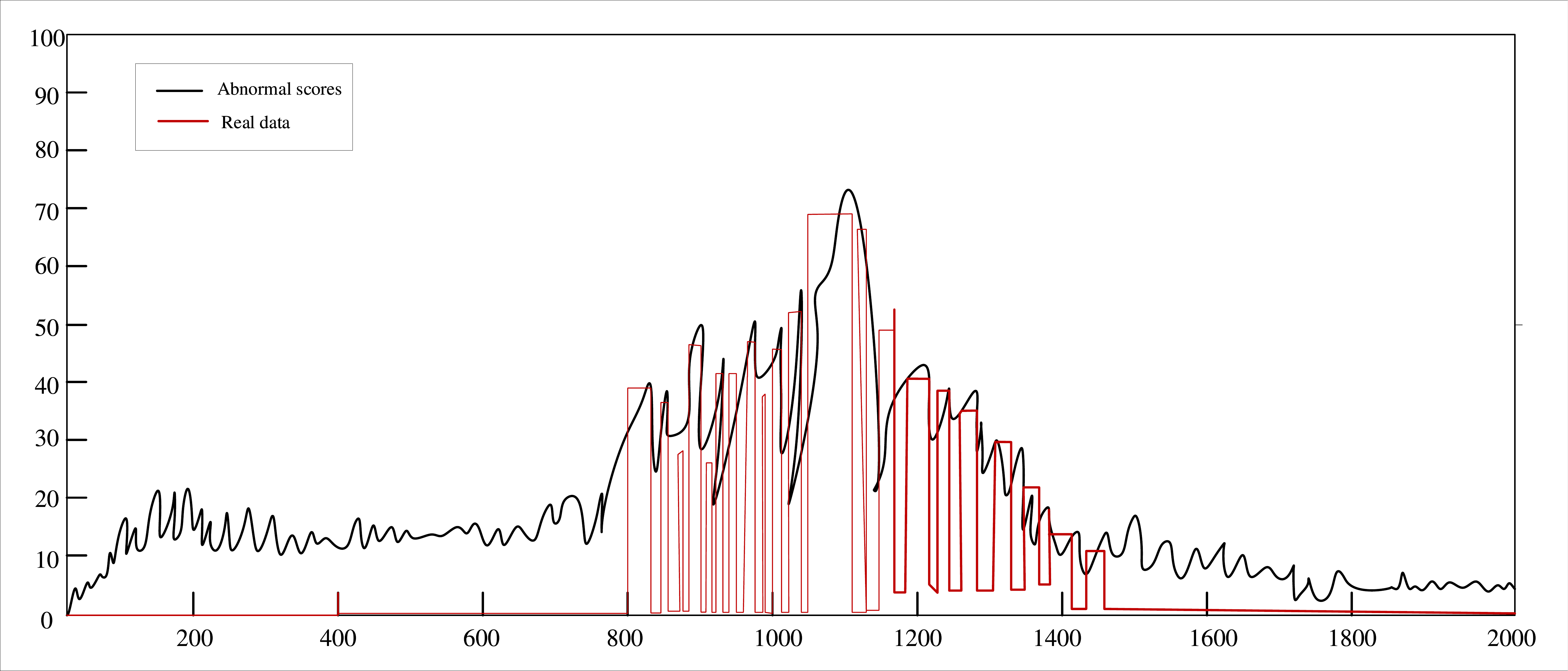}
\caption{Anomaly score obtained by SP-GNN.}
\end{figure}

The clustering results of the proposed method are shown in Fig. 8, and the clustering results of CNN and LSTM are showed in Fig. 9, in Fig. 10 respectively. From the three figure, we can seen our method is superior to the other methods.

\begin{figure*}[h]
    \centering
   \subfigure[The result of SP-GNN.]{
         \includegraphics[width=0.3\textwidth]{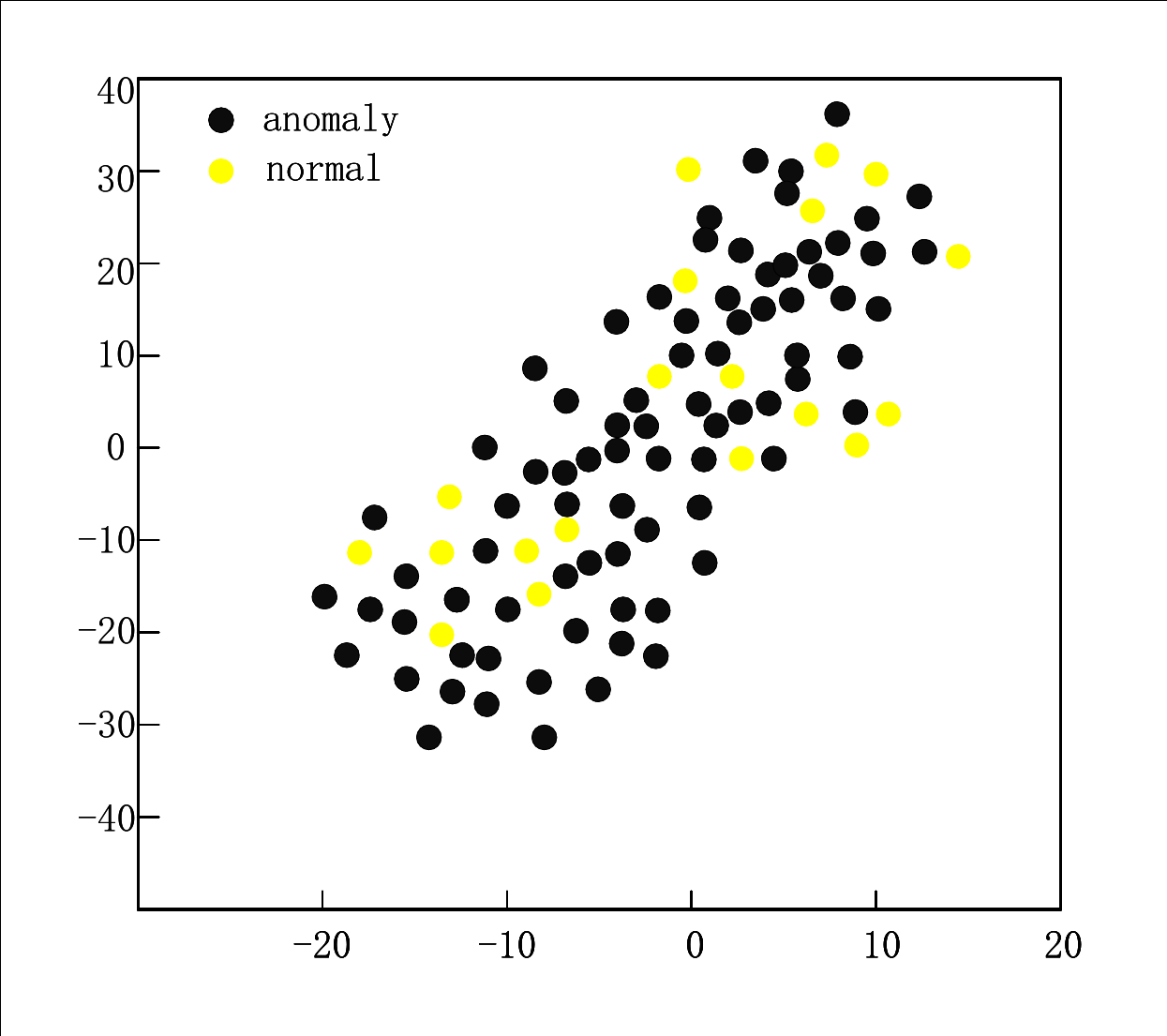}}
    \subfigure[The result of CNN.]{
         \includegraphics[width=0.3\textwidth]{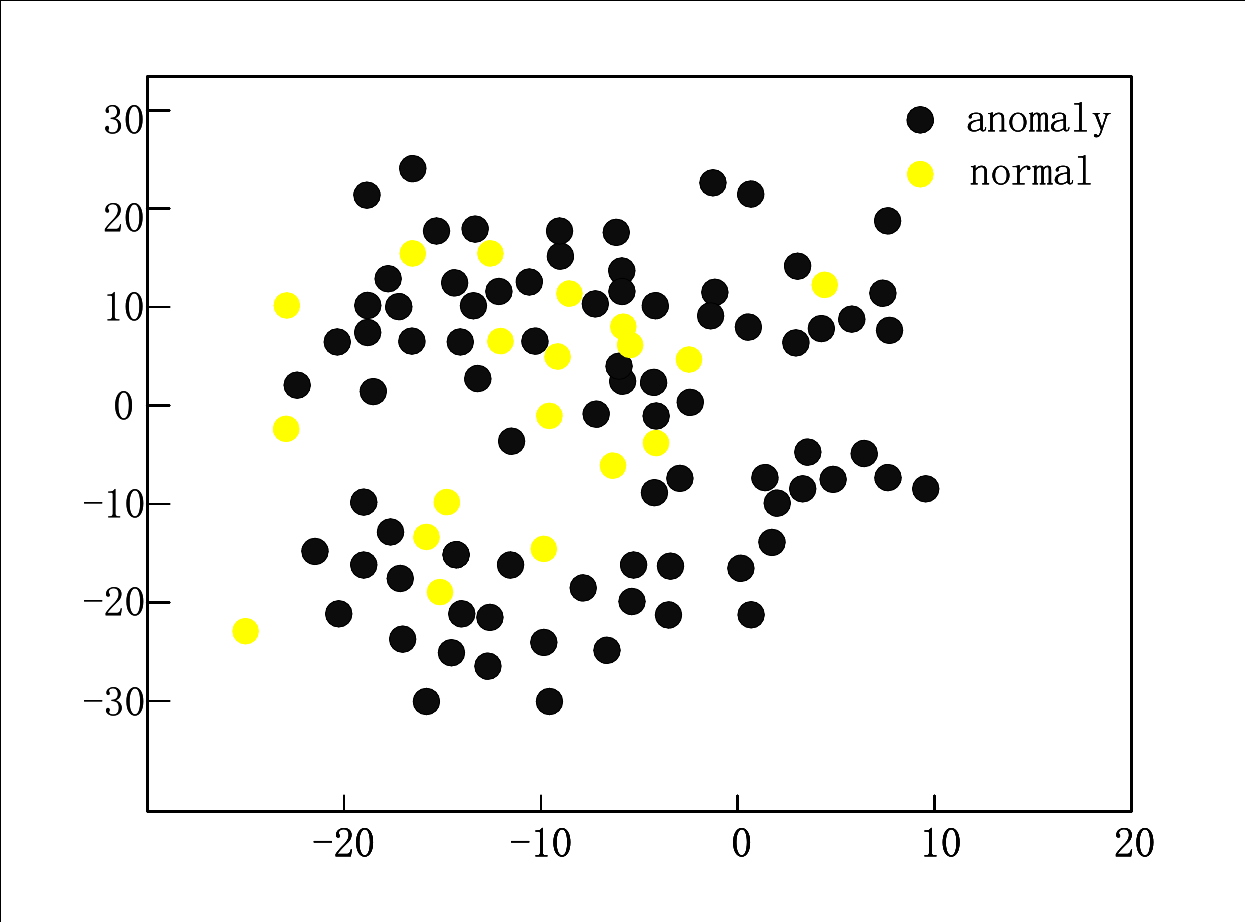}}
    \subfigure[The result of LSTM.]{
         \includegraphics[width=0.3\textwidth]{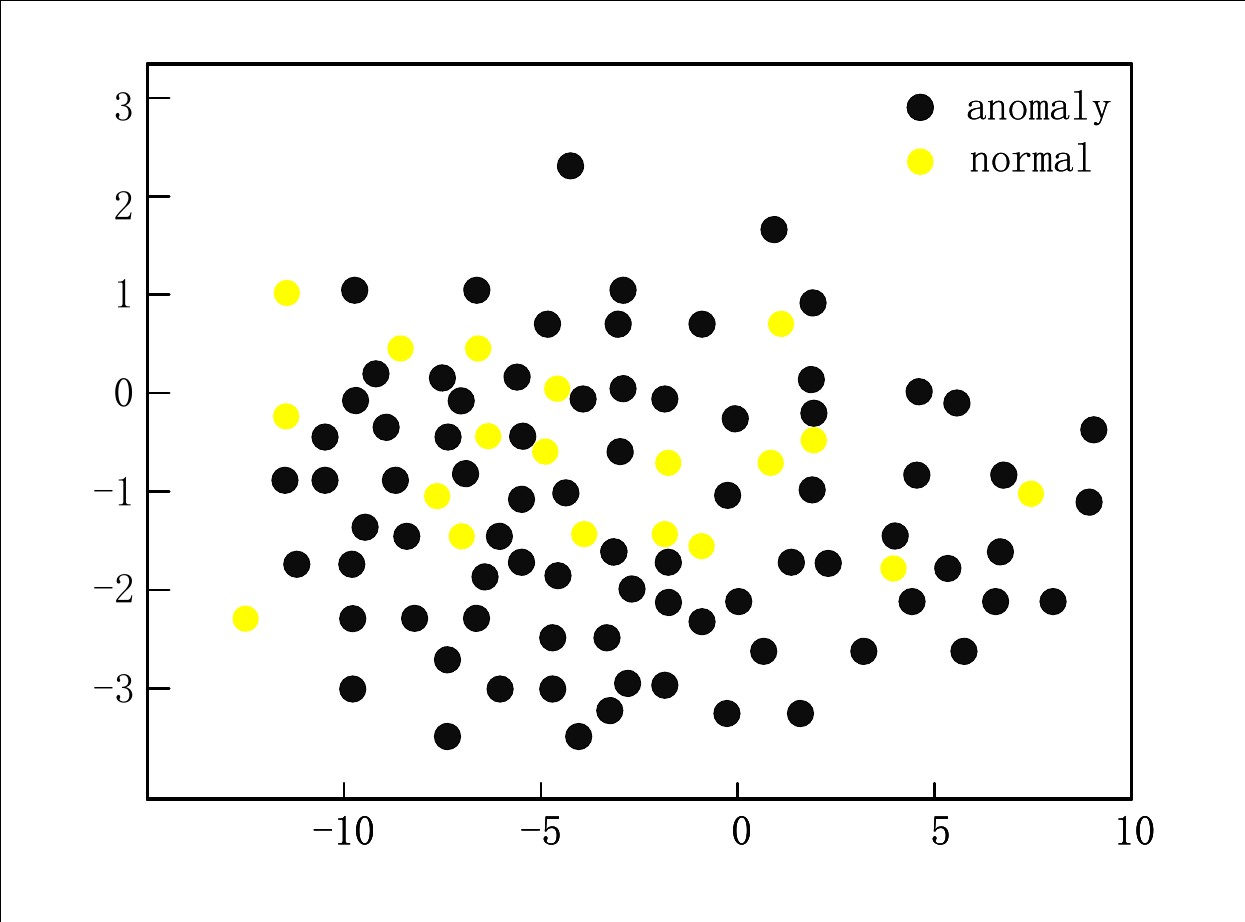}}
        \caption{Experimental Results}
        \label{fig:three graphs}
\end{figure*}

\section{Conclusion}
We address the problem of anomaly event detection in secure critical infrastructure domain.
In the paper, we design a novel anomaly detection method based on multi-source data.
We first leverage spectral clustering algorithm for feature extraction and fusion of multiple data sources. 
Second, we adopt deep graph neural network(Deep-GNN) to perform a fine-gained anomaly social event detection, which reveals the threatening events and guarantee the critical infrastructure security.  Experimental results show that our method has good robustness and adaption. Compared with the other anomaly detection algorithms, our method has a good performance index.

\section*{Acknowledgement}
Xu Zhang is supported by Technological Innovation (2020AAA0108405, 2020AAA0108400) and NSFC U1636123.
The rest authors of this paper were supported by National Key Research and Development Program (2019YFC0850105),
Key Research and Development Project of Hebei Province through grant 20310101D. 

\bibliographystyle{ACM-Reference-Format}
\bibliography{reference}

\end{document}